\documentclass[aps,prl,twocolumn,superscriptaddress,nobibnotes]{revtex4}
\usepackage{graphicx}

\begin{document}

\title{Strongly correlated photons on a chip}

\author{Andreas Reinhard}\thanks{These authors contributed equally to this work.}
\affiliation{Institute of Quantum Electronics, ETH Zurich, 8093
Zurich, Switzerland}
\author{Thomas Volz}\thanks{These authors contributed equally to this work.}
\affiliation{Institute of Quantum Electronics, ETH Zurich, 8093
Zurich, Switzerland}
\author{Martin Winger}
\affiliation{Institute of Quantum Electronics, ETH Zurich, 8093
Zurich, Switzerland}
%\affiliation{Thomas J. Watson, Sr., Laboratory of Applied Physics,
%California Institute of Technology, Pasadena, California 91125, USA}
\author{Antonio Badolato}
\affiliation{Department of Physics and Astronomy, University of
Rochester, Rochester, NY 14627, USA}
\author{Kevin J.\ Hennessy}
\affiliation{Institute of Quantum Electronics, ETH Zurich, 8093
Zurich, Switzerland}
\author{Evelyn L.\ Hu}
\affiliation{School of Engineering and Applied Physics, Harvard
University, Cambridge, Massachusetts 02138, USA}
\author{Ata\c c Imamo\u glu}
\affiliation{Institute of Quantum Electronics, ETH Zurich, 8093
Zurich, Switzerland}

\date{\today}
\maketitle

{\bf Optical non-linearities at the single-photon level are key
ingredients for future photonic quantum technologies
\cite{Obrien:NatPhot09}. Prime candidates for the realization of
strong photon-photon interactions necessary for implementing quantum
information processing tasks \cite{Mabuchi:Science02} as well as for
studying strongly correlated photons
\cite{Hartmann:NatPhys06,Greentree:NatPhys06} in an integrated
photonic device setting are quantum dots embedded in photonic
crystal nanocavities. Here, we report strong quantum correlations
between photons on picosecond timescales. We observe (a) photon
antibunching upon resonant excitation of the lowest-energy polariton
state, proving that the first cavity photon blocks the subsequent
injection events, and (b) photon bunching when the laser field is in
two-photon resonance with the polariton eigenstates of the second
Jaynes-Cummings manifold, demonstrating that two photons at this
color are more likely to be injected into the cavity jointly, than
they would otherwise. Together, these results demonstrate
unprecedented strong single-photon non-linearities, paving the way
for realizing a single-photon transistor \cite{Chang:NatPhys07} or a
quantum optical Josephson interferometer \cite{Gerace:NatPhys09}.}

Cavity quantum electrodynamics (cQED) studies the quantum limit of
light-matter interaction where a single two-level quantum emitter is
coupled to a single cavity mode \cite{Mabuchi:Science02}. In the
strong coupling regime of cavity-QED where the coherent interaction
strength between the emitter and the cavity mode exceeds the
dissipative rates, the elementary excitations (polaritons) have an
anharmonic spectrum (Fig.~1a). As a consequence, this system
embodies the ultimate non-linear optical device enabling the
observation of photon-photon interactions at the single-photon level
\cite{Imamoglu:PRL97}. Various implementations of cavity-QED systems
have been reported with atoms in high finesse cavities
\cite{Brune:PRL96}, with quantum dots (QDs) in different types of
monolithic cavities
\cite{Yoshie:Nature04,Reithmaier:Nature04,Peter:PRL05,Hennessy:Nature07}
and in the microwave domain
\cite{Fink:Nature08,Deppe:NatPhys08,Bishop:NatPhys09}. Recent
experiments in the optical domain using a single atom coupled to
Fabry-Perot \cite{Birnbaum:Nature05} or toroidal cavities
\cite{Dayan:Science08} have demonstrated the photon blockade effect
by observing photon antibunching in correlation measurements. Photon
bunching upon two-photon excitation of the second Jaynes-Cummings
manifold has been observed for a single atom cavity-QED system
\cite{Kubanek:PRL08}. In the solid-state, early results in quantum
dot (QD) cavity-QED systems indicating optical non-linearities have
been reported \cite{Faraon:NatPhys08,Kasprzak:NatMat10}.

\begin{figure}
\includegraphics[width= 86 mm]{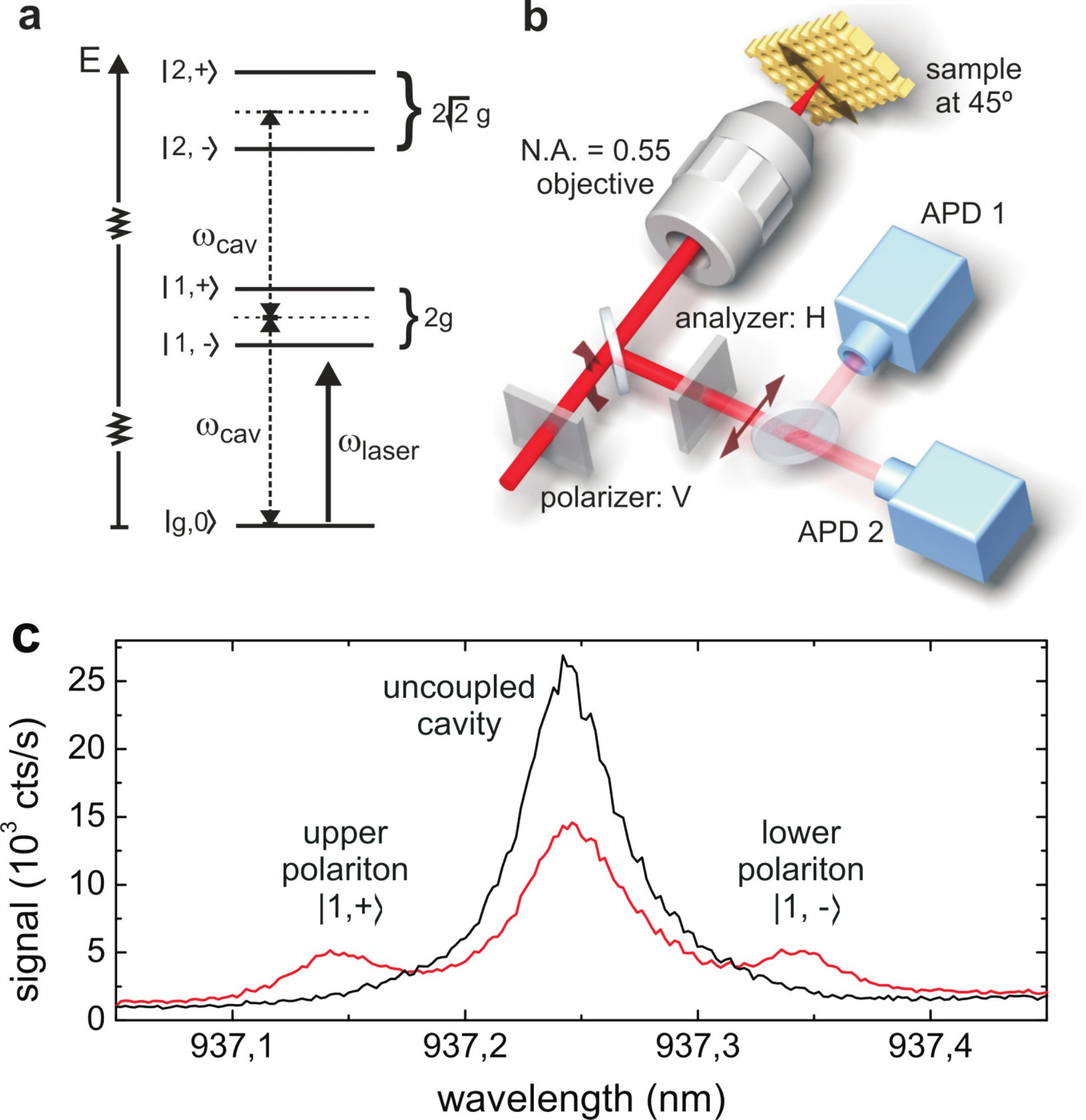}
\caption{ {\bf Resonant scattering spectroscopy of a QD strongly
coupled to a PC cavity.}  a) Non-linear Jaynes-Cummings level scheme
up to the second manifold. b) Sketch of the experimental setup with
crossed-polarized laser suppression. c) On-resonance cw scattering
spectrum for a probe power of $1~\mathrm{nW}$. The black trace was
recorded without the additional re-pump laser. With the re-pump
switched on, the resonant signal from the polaritons is restored
(red trace).} \label{Fig1}
\end{figure}

Here, we show that a single QD deterministically coupled to a
photonic crystal (PC) defect cavity exhibits pronounced photon
antibunching as witnessed by a reduction in two-photon scattering
events from the strongly coupled polariton mode by more than $40~\%$
upon resonant excitation -- a clear indication of photon blockade.
By tuning either the laser or the cavity mode frequency to ensure
resonant two-photon excitation of the higher polariton states
\cite{Schneebeli:PRL08}, we observe that the photon emission is
strongly bunched with an increase in simultaneous two-photon
scattering events by $50~\%$. Hence, we demonstrate for the first
time direct access to the second manifold of the Jaynes-Cummings
ladder in an integrated solid-state photonic device.

By positioning an InAs/GaAs QD at an electric-field antinode of a PC
defect cavity in L3 geometry \cite{Hennessy:Nature07}, we achieve a
coherent coupling constant $g$ of $141~\mathrm{\mu eV}$. The cavity
has a quality factor Q of about $25~000$, corresponding to a cavity
photon decay rate $\kappa$ of $53~\mathrm{\mu eV}$. In order to tune
the cavity frequency, we employ a thin-film deposition technique
using nitrogen gas injection \cite{Strauf:APL06}. A
crossed-polarization technique ensures efficient suppression of the
excitation-laser light which is back-reflected from the sample
surface~\cite{Englund:Nature07,Srinivasan:Nature07}; the setup is
sketched in Fig.~1b.

A large majority of QDs \cite{Santori:PRB04}, as well as most other
solid-state based emitters such as NV centers \cite{Manson:Diam05},
exhibit the phenomenon of blinking, induced by spontaneous or
induced change of the internal state of the QD, and leading to an
intermittent optical response. Experimentally, we find that the
fractional amount of time our system spends in the neutral QD ground
state strongly decreases with increasing resonant probe power, and
is vanishingly small above $1~\mathrm{nW}$ (see Supplementary Material).
Incident laser photons
thus simply scatter off the uncoupled cavity, resulting in a
spectrum as shown in Fig.~1c (black trace). In order to overcome
this problem, we use an off-resonant laser just below the edge of
the wetting layer to re-pump the QD into its neutral ground state.
By alternating re-pump and probe
intervals, we partly recover the polariton spectrum shown in Fig~1c
(red trace).

\begin{figure}
\includegraphics[width= 86 mm]{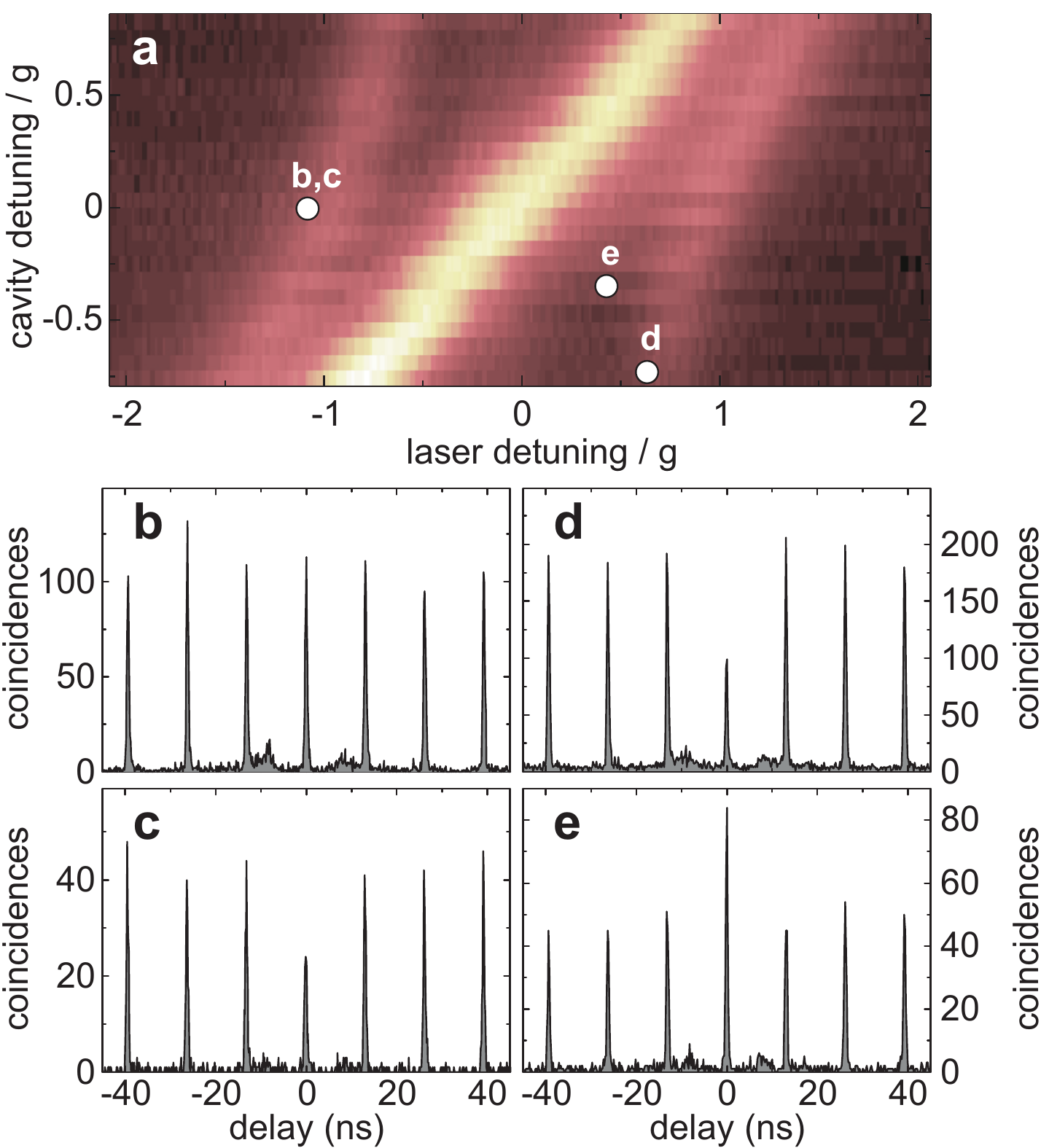}
\caption{ {\bf From Poissonian light to antibunched and bunched
photon streams.} a) Resonant scattering spectra for different cavity
detunings close to the neutral exciton resonance. The circles
indicate the detunings for which the correlation histograms in b --
e were taken. The vertical axes in the histograms specify the photon
coincidences per time-bin of $192~\mathrm{ps}$. b) Autocorrelation
histogram recorded on exact cavity-exciton resonance without re-pump
laser for a detuning corresponding to the energy of the upper
polariton. c) Autocorrelation with the same detunings, but with a
re-pump laser applied. The scattered photons in this case exhibit
photon antibunching. d) Autocorrelation on the lower polariton for a
blue detuning of the cavity of $0.68~g$, where the longer polariton
lifetime ensures stronger photon antibunching. e) Pronounced
photon-bunching on the lower two-photon resonance. Note that the
small features at $\approx\pm 10~\mathrm{ns}$ in b -- e originate
from cross-talk between the APDs due to secondary photon-emission
events. \label{Fig2}}
\end{figure}

We carry out correlation measurements using resonant pulses at a
duration of $T_{\mathrm{pulse}}\approx 72~\mathrm{ps}$ (FWHM),
yielding a ratio between pulse length and polariton lifetime of
$\frac{\kappa}{2}\cdot T_{\mathrm{pulse}}\approx 2.9$. Fig.~2
displays correlation histograms for different laser/cavity
detunings. The respective detunings are indicated in the resonant
scattering spectrum shown in Fig 2a. We define the correlation
function for pulsed excitation at zero time delay
$g_{\mathrm{pulsed}}^{(2)}(0)$ as the area of the central peak
divided by the average area of peaks at other times (see
Supplementary Material). First, we choose the frequency of the
pulsed laser to match the upper polariton transition and obtain
significant antibunching with $g_{\mathrm{pulsed}}^{(2)}(0) = 0.75
\pm 0.06$ (Fig.~2c). For comparison, we turn off the re-pump laser
and verify that photons scattered from the uncoupled cavity mode
have Poissonian statistics (Fig.~2b). As a cross-check, we also
confirmed that the scattered photons from a resonantly driven cavity
mode that is far detuned from all QD transitions exhibit Poissonian
statistics. These experiments jointly demonstrate that the applied
re-pump laser can be used to switch on the single-photon
non-linearity and to control the statistics of the scattered
photons. We emphasize that the degree of photon antibunching
observed in the experiment is limited by the finite excitation pulse
length. Accordingly, a higher degree of antibunching
($g_{\mathrm{pulsed}}^{(2)}(0) = 0.57 \pm 0.02$) is observed when
the cavity is off-resonant from the exciton, as shown in Fig.~2d for
the lower polariton in the case of blue-detuning of the cavity mode;
here the polariton state lifetime is prolonged, making multiple
photon absorption events within a single pulse less likely.
Experimentally, we do not observe a difference between correlations
recorded on the upper and lower polariton if the respective laser
and cavity detunings from the exciton change their sign.

When tuning the laser photon energy to half the energy of the lower
polariton eigenstate of the second Jaynes-Cummings manifold, we
observe photon-bunching. Fig.~2e displays a corresponding
correlation histogram with $g_{\mathrm{pulsed}}^{(2)}(0) = 1.5 \pm 0.1$.
The bunching has its origin in a two-photon transition from
the ground state to the second manifold of the Jaynes-Cummings
ladder. We emphasize that the non-vanishing correlations at zero
time delay in the case of photon blockade, and the moderate
bunching-feature on the two-photon transition have their origin
mostly in the particular implementation of the measurement using
pulsed laser excitation and slow single-photon detectors; in order
to confirm that and to explain the principal experimental features,
we carried out numerical simulations of
$g_{\mathrm{pulsed}}^{(2)}(0)$ using a Monte Carlo wave function
(MCWF) approach (see Supplementary Material). Here, we directly
accounted for the pulsed laser excitation, the non-zero laser
background due to imperfect extinction of the laser reflection and
the uncoupled cavity resonance due to the blinking of the dot.
Fig.~3a displays the resulting
autocorrelation function $g_{\mathrm{pulsed}}^{(2)}(0)$ at zero time
delay for varying cavity and laser detunings.

In addition to the bunching features originating from two-photon
resonances, we would normally expect strong bunching when the
external laser that drives the cavity mode is resonant with the bare
exciton \cite{Kubanek:PRL08}, as can be seen in Fig.~3a (dotted
vertical line). Unlike in previous experiments
\cite{Faraon:NatPhys08}, this bunching-feature is absent in our
experiments as well as in the simulations when the cavity frequency
is resonant with the bare QD exciton, since for this detuning we
predominantly detect photons from instances when the QD remains
charged and is thus off-resonant. Conversely, our experiments
demonstrate for the first time that a coupled QD-cavity device
realizes the anharmonic Jaynes-Cummings model \cite{footnote:1}. In
the simulations, we assumed a pure QD dephasing rate of
$\hbar\gamma_{deph} = 13~\mathrm{\mu eV}$ in addition to the cavity
dissipation rate $\kappa$, consistent with the polariton linewidths
observed in the experiment. This line broadening might stem from
charge fluctuations in the QD environment, partially induced by the
re-pump laser, and from phonon-induced dephasing.

\begin{figure}
\includegraphics[width= 86 mm]{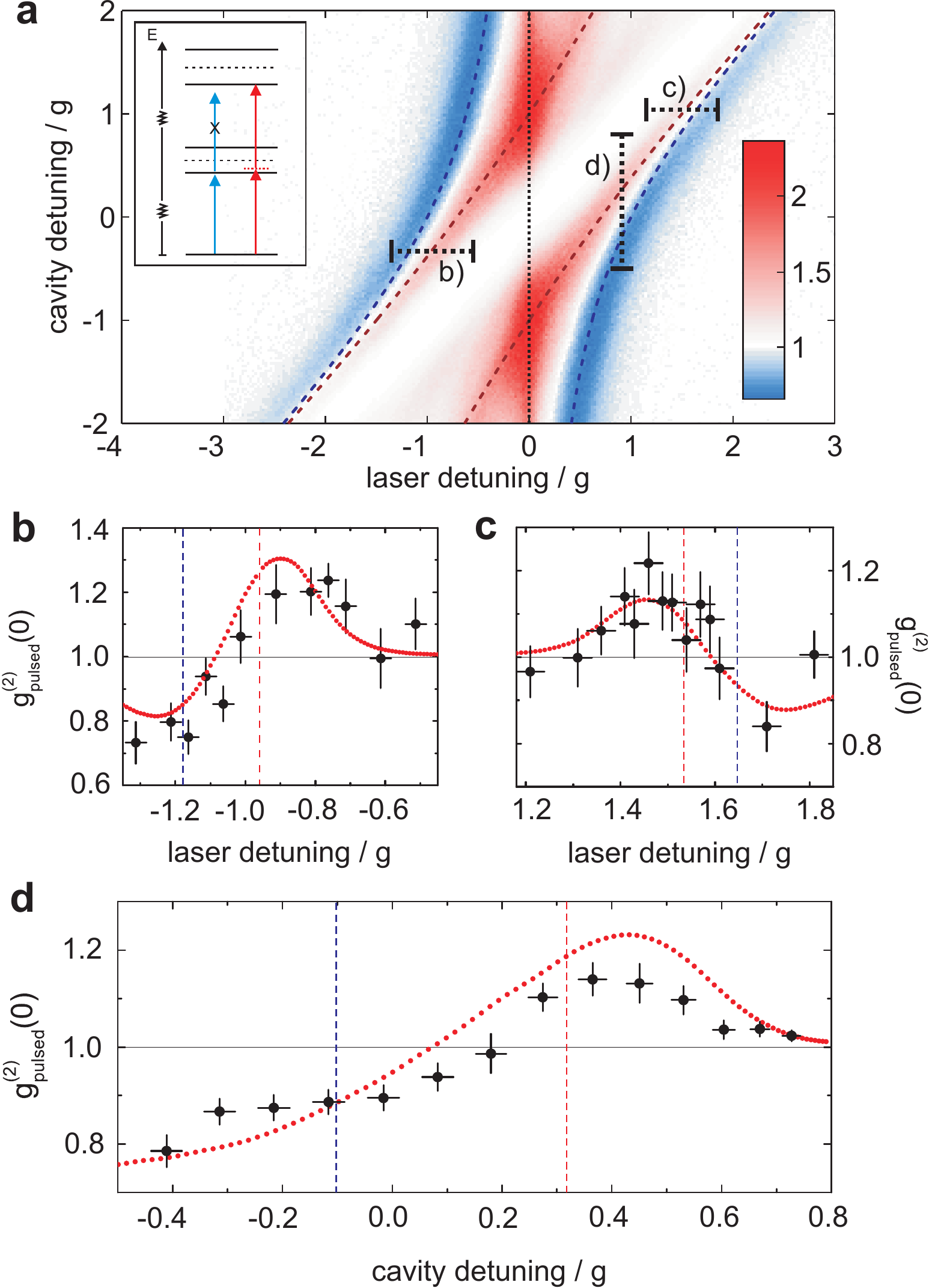}
\caption{{\bf Calculated and measured autocorrelation functions
$g_{\mathrm{pulsed}}^{(2)}(0)$ for different cavity and laser
detunings.} a) Results of a MCWF simulation close to the neutral
exciton-cavity resonance using experimental parameters determined
from linear resonant light scattering measurements. The blue regions
correspond to detunings with sub-Poissonian statistics, whereas the
red regions indicate photon-bunching. Blue and red dashed lines
denote the expected positions of the polariton frequencies and the
two-photon resonances, respectively. The enhanced bunching feature
along the dotted vertical line at zero detuning is a consequence of
an interference effect, ensuring small occupancy of the cavity mode.
The bold dotted lines indicate the parameter ranges of the three
different data traces displayed in b -- d. There, the red dotted
traces are the theoretical expectations from the MCWF simulations.}
\label{Fig3}
\end{figure}

In Fig.~3b and 3c we probe the upper and lower polariton branches
for constant cavity detunings of $\left(-0.33 \pm 0.04\right)g$ and
$\left(1.04 \pm 0.03\right)g$, respectively, by tuning the laser
wavelength. The continuous change in photon correlations from strong
antibunching to bunching maps out the (anharmonic) spectrum of the
Jaynes-Cummings ladder. In addition, we perform correlation
measurements for a varying cavity-exciton detuning at a constant
laser-exciton detuning of $\left(0.94 \pm 0.05\right)g$,
demonstrating that the nature of strong photon correlations can be
tuned by changing either the laser or the cavity mode frequency. The
agreement between experimental values and theoretical expectation is
very good in all cases; we emphasize in this regard that the
theoretical curves provide a prediction based on independently
determined parameters. The dashed vertical lines in Fig.~3b-d
correspond to the polariton energy (blue) and the energy of the two-photon
transition to the second manifold (red), indicating the origin of both the
antibunching and the bunching features.

An obvious extension of the results presented here includes
non-linear optics experiments at the single-photon level, e.g. the
realization of a single-photon transistor \cite{Chang:NatPhys07}.
Furthermore, our results elevate this system to an ultimate optical
non-linear building block for more complex structures investigating
strong photon correlations in non-equilibrium settings, such as an
optical Josephson interferometer \cite{Gerace:NatPhys09} or coupled
arrays of non-linear cavities
\cite{Hartmann:NatPhys06,Greentree:NatPhys06,Carusotto:PRL09}. In
connection with recent progress on the fabrication of
site-controlled QDs within arrays of photonic crystal cavities
\cite{Gallo:APL08} and the ability to tune QD transitions by up to
$25~\mathrm{meV}$ in p-i-n structures \cite{Bennett:APL10}, the present work
demonstrates the great potential of QD-cavity systems as candidates
for photonic quantum simulators.

This work is supported by NCCR Quantum Photonics (NCCR QP), research
instrument of the Swiss National Science Foundation (SNSF) and an
ERC Advanced Investigator Grant (A.I.). The authors thank Iacopo
Carusotto for helpful discussions. The authors declare that they
have no competing financial interests. Correspondence and requests
for materials should be addressed to T.V. and A.I. (E-mail:
volz@phys.ethz.ch, imamoglu@phys.ethz.ch)

\section*{Methods}

\subsection*{Optical characterization}
Optical characterization is performed in a liquid helium flow
cryostat at a temperature $T=5-7~\mathrm{K}$. A $50 \times$
microscope objective ($\mathrm{NA}=0.55$) is used to illuminate the
defect region of the PC with a spot size of about $1~\mathrm{\mu
m}$. Emitted photons are collected with the same objective. After
cooldown, the cavity luminesces at a wavelength of around
$935~\mathrm{nm}$, as determined with photoluminescence
spectroscopy. In order to tune the cavity into resonance with the
neutral exciton transition, a thin-film deposition technique using
nitrogen gas injection is used. Even without active deposition of
nitrogen, the cavity mode shows an intrinsic red tuning at a rate up
to $0.008~\mathrm{nm/hour}$. For resonant scattering experiments, we
excite the cavity mode with an actively power-stabilized narrow
bandwidth, mode-hop-free tunable diode laser. The sample is mounted
such that there is a $45^{\circ}$ angle between the polarization
axes of the cavity mode and the laser. By sending the collected
light through a single-mode fiber with mounted fiber-polarization
controllers and a subsequent analyzer, we have precise control over
the extinction of the reflected laser light. The photons scattered
off the cavity are detected with an avalanche photo diode (APD) in
single-photon counting mode.

\subsection*{Off-resonant re-pump scheme}
For the off-resonant re-pumping of the system, we use an additional
cw titanium-sapphire laser  at a wavelength of $857.2~\mathrm{nm}$.
Using an acousto-optical modulator (AOM) we excite the QD with
rectangular laser pulses at a repetition rate of
$0.5-1~\mathrm{MHz}$. In between the re-pump pulses, we probe the
system by triggering the APD readout.

\subsection*{Autocorrelation measurements}
For photon correlation measurements, we replace the cw probe light
by resonant pulses from a mode-locked titanium-sapphire laser in ps
mode, with a repetition rate of $76.3~\mathrm{MHz}$. In order to
filter the broadband pulses we send them through a 750 mm grating
spectrograph and couple the diffracted light into a single-mode
fiber. The transmitted pulse is approximately Gaussian with a
typical width of $0.018~\mathrm{nm}$. Using a streak camera we
confirm that the resulting pulses have near-Fourier-limited
durations of about $72~\mathrm{ps}$. The dependence of photon
correlations on the laser frequency are obtained by rotating the
grating of the spectrograph and consequently filtering out different
parts of the pulse spectrum. Measurements depicting the dependence
of photon correlations on the cavity resonance frequency on the
other hand are carried out by exploiting an intrinsic cavity
wavelength drift of about $0.006~\mathrm{nm/h}$ while keeping the
laser frequency fixed. We perform autocorrelation measurements with
a Hanbury-Brown and Twiss set-up, consisting of a 50:50 beam
splitter and an APD at each output. Photon arrival time differences
are determined with a time-to-amplitude converter and plotted in a
histogram.

\clearpage

\newpage

\section*{Supplementary Material}

\renewcommand{\figurename}{FIG. S}

\section*{Laser-induced quantum dot blinking}

In resonant spectroscopy, far below saturation $p_{\mathrm{sat}}$ of
the system, the detection rate of scattered photons is expected to
grow linearly with excitation power $p_{\mathrm{exc}}$. In stark
contrast, we observe highly nonlinear power dependencies at picoWatt
(pW) level, which is five orders of magnitude smaller than the cw
saturation power $p_{\mathrm{sat}}\approx 850~\mathrm{nW}$ (for an
estimated $2~\%$ laser-cavity coupling). The polariton signal
saturates at $p_{\mathrm{exc}} \approx 10~\mathrm{pW}$, whereas the
uncoupled cavity resonance appears at the same power level with a
super-linear power dependence, as seen in Fig. S~4a. We attribute
this behaviour to laser-induced QD-blinking, which could originate
from relaxation of optically excited QD states either into a dark
exciton or an electron (hole) charged state. The resulting
long-lived states (denoted by $\left|\mathrm{h}\right>$ in Fig. S~4b
shift the exciton resonance off the cavity frequency, such that the
cavity resonance becomes uncoupled from QD resonances. For
excitation powers below $5~\mathrm{pW}$, only the polaritons appear
in the spectrum. For increasing incident optical power, however, the
laser-induced charge/spin-pumping causes a reduction of neutral
ground state occupation such that the polariton signal saturates.
Above $p_{\mathrm{exc}} \approx 100~\mathrm{pW}$, the bare cavity
resonance completely dominates the spectrum.

We determine the average lifetime of these long-lived charged/dark
QD states as well as the neutral ground state lifetime, by measuring
the second order auto-correlation function of the uncoupled cavity
and the polaritons, respectively. By extracting the lifetimes from
the correlation traces, like the ones shown in Fig. S~5a, we
identify the metastable charged/dark QD state lifetimes to be of
several $\mathrm{ms}$. In contrast, the neutral ground state
lifetime is about $\approx 300~\mathrm{\mu s}$ for $p_{\mathrm{exc}}
= 10~\mathrm{pW}$, with a significant reduction for higher
excitation powers, as shown in Fig. S~5b \cite{footnote:S1}. This is
consistent with our assumption and motivates the use of a
non-resonant re-pump laser that randomly re-shuffles the internal QD
state. We implement this strategy by exciting the system just below
the wetting-layer resonance and observe a relaxation probability
into the neutral ground state of $0.3-0.4$. Using a pump-probe
scheme, we are able to partially suppress the influence of blinking
and obtain a time-integrated spectrum that exhibits a three-peak
feature (see Fig.~1c in the main text).

\begin{figure}
\includegraphics[width= 86 mm]{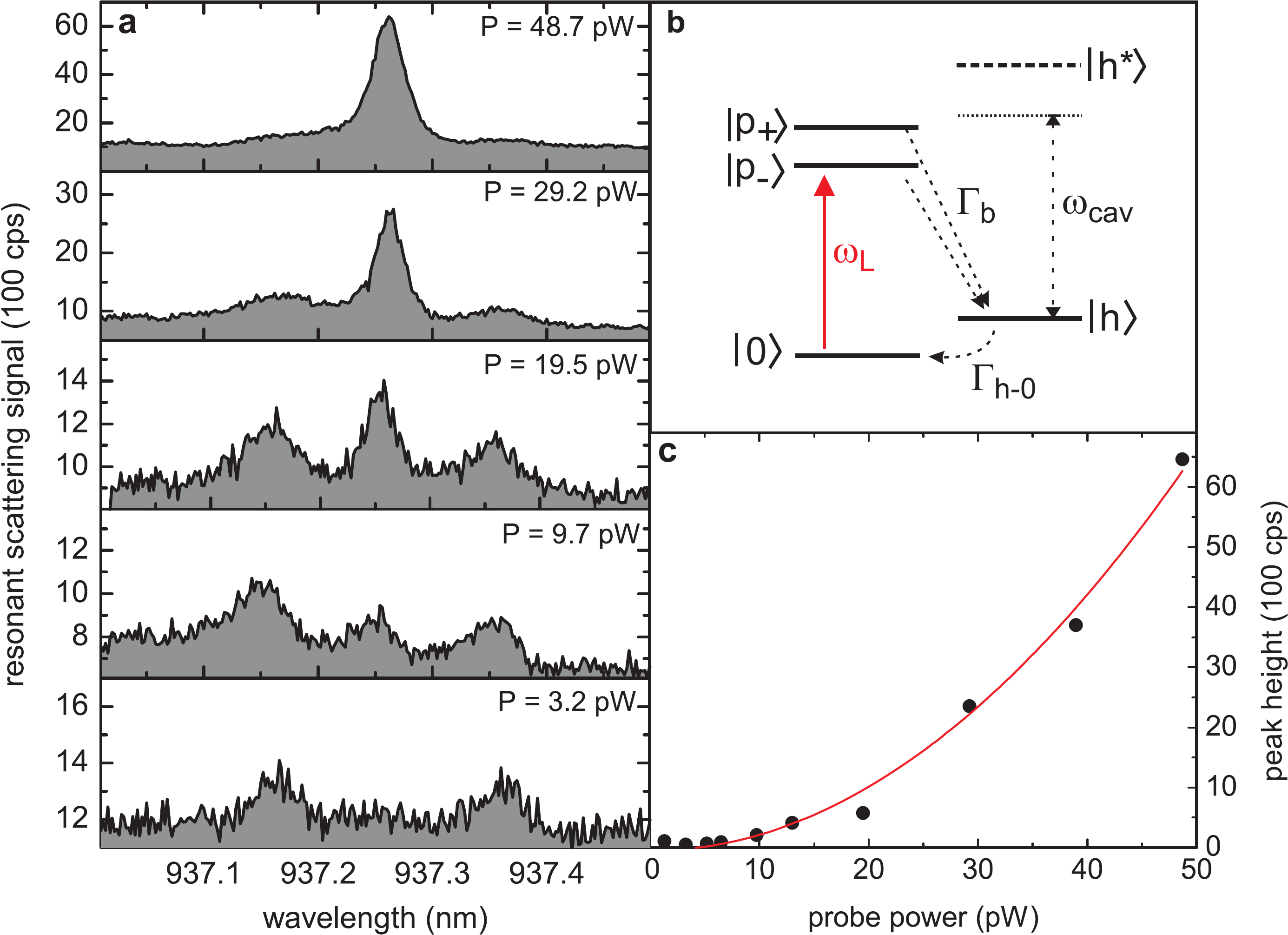}
\caption{ {\bf Low power spectra on cavity-exciton resonance.} a) At
a resonant probe power of $3.2~\mathrm{pW}$ only the polaritons are
observed. For increasing excitation powers, the uncoupled cavity
mode dominates the spectrum, while the polariton peaks saturate. b)
Level scheme: pumping of charged or dark neutral states
$\left|\mathrm{h}\right>$ occur when the polaritons are excited. The
recovery rate into the neutral ground state $\Gamma_{\mathrm{h-0}}$
is below $10^3$ $s^{-1}$. c) Power dependent peak height of the
uncoupled cavity resonance, which provides a direct measure of
charged/dark state occupation.} \label{S1}
\end{figure}

Finally, we emphasize that the blinking observed in our QD-cavity
system is by no means unique and is observed by most groups working
with such systems \cite{Faraon:PhD,Cassabois:Comm}. Depending on the
particular device, however, the effect is more or less pronounced
\cite{Faraon:NatPhys08_S}.

\begin{figure}
\includegraphics[width= 86 mm]{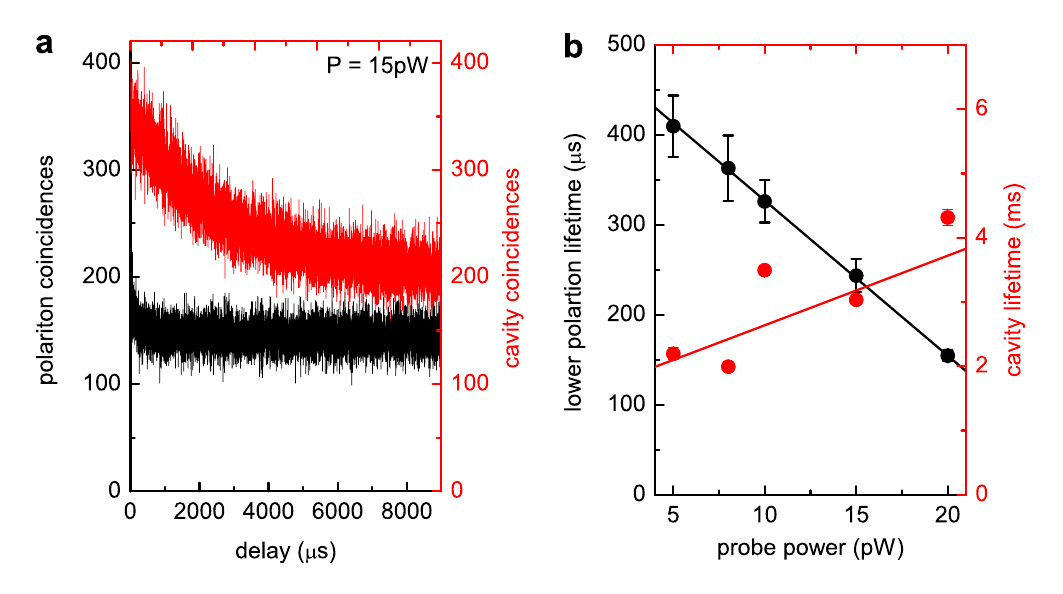}
\caption{ {\bf Lifetimes of the neutral quantum dot ground state and
the charged/dark states.} a) Auto-correlation traces on the lower
polariton resonance (black trace) and the uncoupled cavity resonance
(red trace) at a probe power of $15~\mathrm{pW}$. The lifetime ratio
reveals that the system spends an order of magnitude more time in a
state other than the neutral ground state due to laser-induced
charge/spin pumping. b) Lifetimes of the neutral ground state (black
trace) and the charged/dark states (red trace) as a function of
excitation power.} \label{S2}
\end{figure}

\section*{Determination of second order correlations at zero time delay}

The second order auto-correlation function $g^{\left( 2
\right)}\left(\tau\right)$ provides a direct measure of the
statistics of emitted light. The degree of antibunching or bunching
is simply given by the value at zero time delay $g^{\left( 2
\right)}\left(0\right)$. When the cavity is resonant with the
exciton, we expect $g^{\left( 2 \right)}\left(0\right) = 0.34$ on
the lower polariton according to a simulation using our experimental
parameters. The width of the dip / peak in $g^{\left( 2
\right)}\left(\tau\right)$ is given by the polariton state lifetime,
which is in our case on the order of tens of ps. Since the APDs we
use have a much higher timing jitter, they cannot be used to measure
$g^{\left( 2 \right)}\left(0\right)$ under cw laser excitation. In
order to overcome this limitation, we measure the auto-correlation
function by exciting the system with short resonant pulses with a
pulse-width similar to the polariton lifetimes. We define $g^{\left(
2 \right)}_{\mathrm{pulsed}}\left(0\right)$ as the area of the
central peak divided by the average area of peaks at other times. In
order to have a realistic comparison between theory and our
experimental data, we use a Monte Carlo wave function (MCWF)
approach to  obtain an estimate of $g^{\left( 2
\right)}_{\mathrm{pulsed}}\left(0\right)$.

\section*{A theoretical estimate of $g^{(2)}_{\mathrm{pulsed}}(0)$}

The $g^{\left( 2 \right)}_{\mathrm{pulsed}}\left(0\right)$ we
measure is given by
\begin{equation}
\label{eq:g2pulsed} g^{\left( 2
\right)}_{\mathrm{pulsed}}\left(0\right)=\frac{2\int\limits_{-T_{\mathrm{rep}}/2}^{T_{\mathrm{rep}}/2}dt_{1}\int\limits_{t_{1}}^{T_{\mathrm{rep}}/2}dt_{2}~G^{\left(
2
\right)}\left(t_{1},t_{2}\right)}{\int\limits_{-T_{\mathrm{rep}}/2}^{T_{\mathrm{rep}}/2}dt_{1}\int\limits_{-T_{\mathrm{rep}}/2}^{T_{\mathrm{rep}}/2}dt_{2}~I\left(t_{1}\right)I\left(t_{2}\right)},
\end{equation}
where $G^{\left( 2
\right)}\left(t_{1},t_{2}\right)=\left<\mathrm{g}\right|\hat{C}^{\dagger}\left(t_{1}\right)\hat{C}^{\dagger}\left(t_{2}\right)\hat{C}\left(t_{2}\right)\hat{C}\left(t_{1}\right)\left|\mathrm{g}\right>$,
$I\left(t\right)=\left<\mathrm{g}\right|\hat{C}^{\dagger}\left(t\right)\hat{C}\left(t\right)\left|\mathrm{g}\right>$,
and $T_{\mathrm{rep}}$ is the time difference between two laser
pulses. Here, $\left|\mathrm{g}\right>$ denotes the system ground
state, before a laser pulse has excited the system. The system
collapse operator
$\hat{C}\left(t\right)=\sqrt{\kappa}~\hat{a}\left(t\right)$ evolves
according to the non-Hermitian effective Hamiltonian (in the
interaction picture with the rotating frame of the  laser center
frequency)
$H_{\mathrm{eff}}\left(t\right)=H_{\mathrm{JC}}+H_{\mathrm{int}}\left(t\right)-\frac{i
\hbar}{2}\left(\kappa~\hat{a}^{\dagger}\hat{a}+\gamma~\hat{\sigma}_{+}\hat{\sigma}_{-}+\gamma_{\mathrm{deph}}\right)$,
where $\hat{a}$ is the cavity photon annihilation operator and
$\hat{\sigma}_{+},\hat{\sigma}_{-}$ are the exciton creation and
annihilation operators.
%$\hat{\sigma}_{z}=\hat{\sigma}_{+}\hat{\sigma}_{-}-\hat{\sigma}_{-}\hat{\sigma}_{+}$
$H_{\mathrm{JC}}$ is the Jaynes-Cummings Hamiltonian and
$H_{\mathrm{int}}=\frac{\hbar\Omega\left(t\right)}{2}\left(\hat{a}+\hat{a}^{\dagger}\right)$
denotes the interaction with the Gaussian laser pulse,
$\Omega\left(t\right)=\Omega_{0}~\mathrm{exp}\left(-2~\mathrm{ln}\left(2\right)t^{2}/~T_{\mathrm{pulse}}^{2}\right)$.
$\kappa$~denotes the cavity dissipation rate, $\gamma$ the exciton
spontaneous recombination rate and $\gamma_{\mathrm{deph}}$ the
exciton pure dephasing rate; the latter may stem from phonon
mediated coupling of the two dressed polariton states. In order to
determine $G^{\left( 2 \right)}\left(t_{1},t_{2}\right)$ and
$I\left(t\right)$, we use a Monte Carlo wavefunction approach, as
described in Ref.~\cite{Molmer:OptSoc93}. The calculations are
performed with the experimentally determined values,
$\lambda_{\mathrm{exciton}}=937.25~\mathrm{nm}$, $\hbar
g=141~\mathrm{\mu eV}$, $\hbar\kappa=53~\mathrm{\mu eV}$,
$\hbar\gamma=0.66~\mathrm{\mu eV}$, and
$\hbar\gamma_{\mathrm{deph}}=13~\mathrm{\mu eV}$. For a typical
pulse duration of $72~\mathrm{ps}$ (FWHM), a pulse energy of
$82~\mathrm{eV}$ and an estimated laser -- cavity coupling
efficiency of $2~\%$, we obtain $\hbar\Omega_{0}\approx
24~\mathrm{\mu eV}$.

In the experiments, the suppression of the non-interacting reflected
laser light is not perfect $(\approx~1:50~000)$, and consequently,
our measurements are subject to interference effects between cavity
emission and laser light. Possible reasons for incomplete
suppression are a mixing of polarization states when the light
traverses the collection fiber or the analyzer, or depolarization
effects due to the reflection at the photonic crystal membrane. We
take the phase difference $\phi$ between the laser light that drives
the Jaynes-Cummings system and the non-interacting (reflected) laser
to be $\phi=-\pi/2$; this value corresponds to the (experimentally
optimized) largest ratio between detected cavity photons and laser
background. We emphasize that we have a control over both the amount
of laser suppression and the phase difference $\phi$, by the use of
fiber-polarization controllers.

As described in the previous section, the system is in the desired
neutral ground state in only $r=30-40~\%$ of the cases in average.
Experimentally, we cannot distinguish between on- and off-resonant
cases, and the resulting correlation function is a sum of both
on-resonant and the Poissonian off-resonant correlation functions.
Thus, at zero time delay
\begin{equation}
\label{eq:g2tot}
\tilde{g}^{\left(2\right)}_{\mathrm{pulsed}}\left(0\right)=\frac{r
E_{\mathrm{on}}^{2}
g^{\left(2\right)}_{\mathrm{pulsed}}\left(0\right)+\left(1-r\right)
E_{\mathrm{off}}^{2}}{r E_{\mathrm{on}}^{2}+\left(1-r\right)
E_{\mathrm{off}}^{2}},
\end{equation}
where $E_{\mathrm{on}}$ is the average detected energy per pulse of
the Jaynes-Cummings system, as described above, and
$E_{\mathrm{off}}$ the respective pulse energy of the uncoupled
cavity emission, together with the non-interacting laser light
background. Close to the cavity-exciton resonance,
$g^{(2)}_{\mathrm{pulsed}}(0)$ should exhibit a bunching peak on the
exciton transition (see Fig. S~6a). This bunching-feature is absent
in $\tilde{g}^{(2)}_{\mathrm{pulsed}}(0)$ (Fig. S~6b) due to the
predominant detection of photons scattered from the uncoupled cavity
mode. For the rest of the discussion here and in the main text, we
will omit the tilde and refer to equation (\ref{eq:g2tot}) as the
definition of correlations at zero time delay.

\begin{figure}
\includegraphics[width= 86 mm]{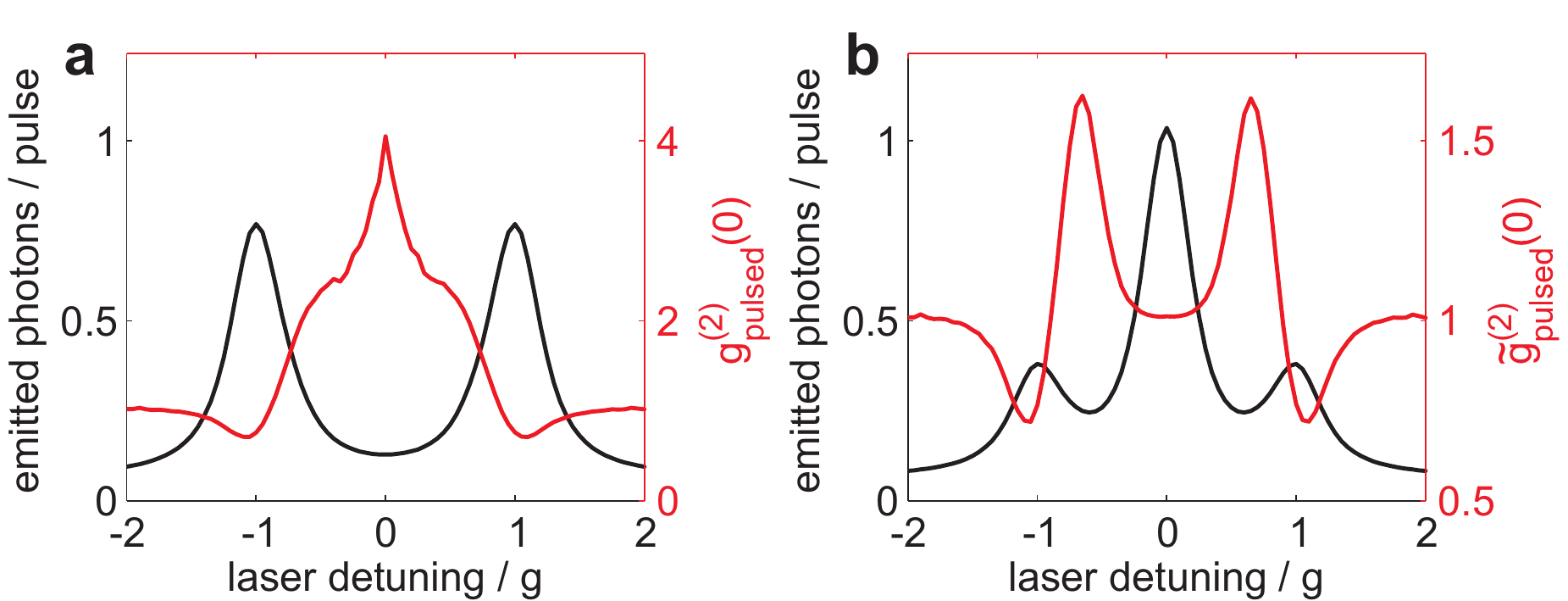}
\caption{ {\bf MCWF simulations on cavity-exciton resonance.} a)
Assuming that there is no QD-blinking, the system spectrum exhibits
only the two polariton peaks (black line). There is a strong
bunching peak for zero laser-exciton detuning in
$g^{(2)}_{\mathrm{pulsed}}(0)$ (red line). b) Simulations of the
real system with QD-blinking ($r=0.4$) yield a spectrum (in time
average) with a three-peak feature (black line); the middle
uncoupled cavity peak causes the bunching feature in
$\tilde{g}^{(2)}_{\mathrm{pulsed}}(0)$ to vanish (red line).}
\label{S3}
\end{figure}

\section*{Pulse parameters}

Simulations on the lower polariton reveal a reduction of the
non-classical part $\left|g^{\left( 2
\right)}_{\mathrm{pulsed}}\left(0\right)-1\right|$. The reasons are
repeated photon absorption/emission events within a single laser
pulse, excitations of more than one resonance due to the broad pulse
spectrum and the large laser-cavity coupling with $\Omega_{0}
\lesssim \kappa$. For Fourier-limited Gaussian laser pulses, we find
a minimum of $g^{\left( 2
\right)}_{\mathrm{pulsed}}\left(0\right)=0.75$ for a laser pulse
duration of about $50~\mathrm{ps}$, as shown in Fig. S~7a. In the
experiments, we measure the spectrally filtered pulses to have
slightly larger durations of $T_{\mathrm{pulse}}\approx
72~\mathrm{ps}$. This corresponds to a ratio between pulse duration
and polariton lifetime of $\frac{\kappa}{2}\cdot
T_{\mathrm{pulse}}\approx 2.9$. For increasing pulse energy, the
calculated $\left|g^{\left( 2
\right)}_{\mathrm{pulsed}}\left(0\right)-1\right|$ decreases, as
shown in Fig. S~7b. We typically work with pulse energies between
$40$ and $100~\mathrm{eV}$, which corresponds to $0.6 - 1.5$
scattered photons per pulse on laser-cavity resonance for the
estimated coupling efficiency of $2~\%$.

\begin{figure}
\includegraphics[width= 86 mm]{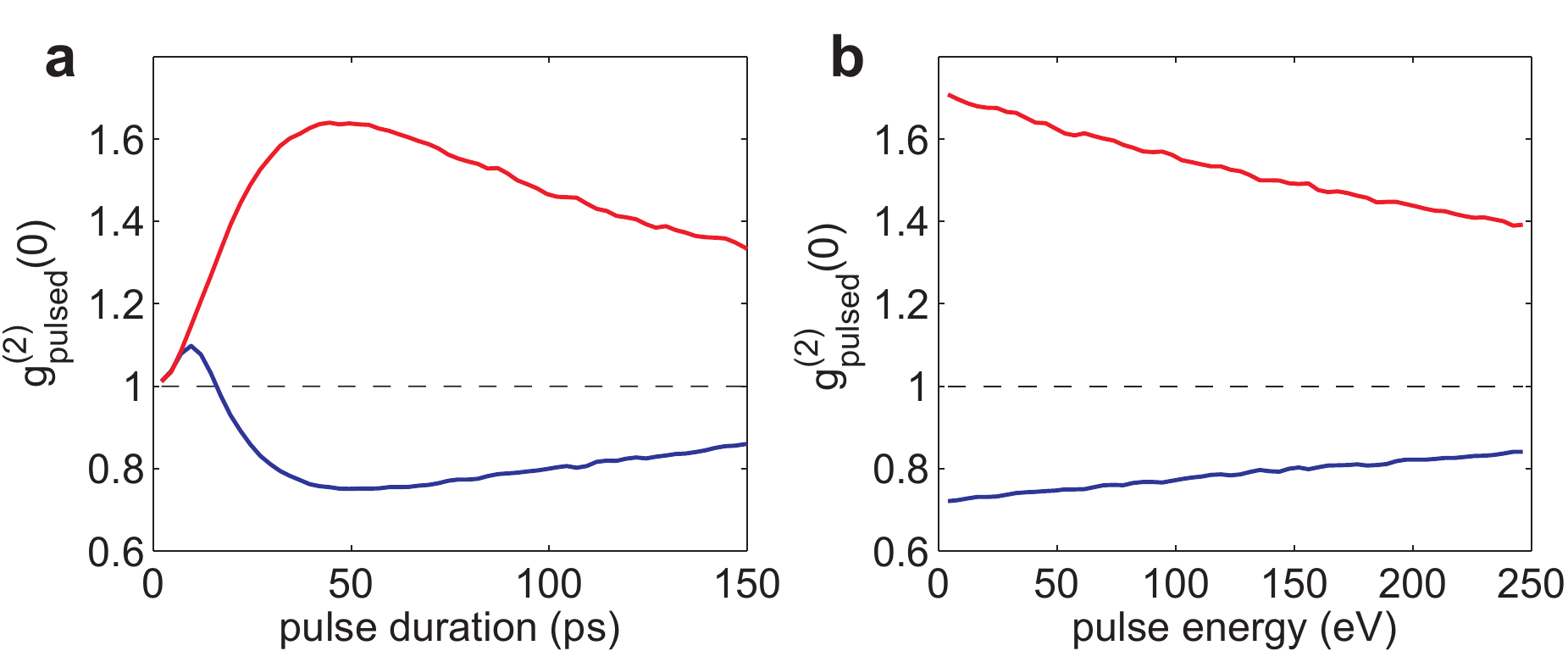}
\caption{ {\bf Simulations of $g^{(2)}_{\mathrm{pulsed}}(0)$ for
different pulse parameters.} Blue traces indicate probing of the
lower polariton and red traces the lower two-photon resonance. a)
$g^{\left( 2 \right)}_{\mathrm{pulsed}}\left(0\right)$ as a function
of pulse duration for a fixed pulse energy of $82~\mathrm{eV}$. b)
$g^{\left( 2 \right)}_{\mathrm{pulsed}}\left(0\right)$ as a function
of pulse energy for $72~\mathrm{ps}$ pulses.} \label{S4}
\end{figure}

The efficiency of the generation of non-classical light improves
with a slight cavity-exciton detuning. A blue detuning of the cavity
by $g$ with respect to the exciton resonance yields $g^{\left( 2
\right)}_{\mathrm{pulsed}}\left(0\right)=0.63$ for the lower
polariton. This reduction in $g^{\left( 2
\right)}_{\mathrm{pulsed}}\left(0\right)$ originates predominantly
from the prolongation of the polariton lifetime which reduces the
likelihood of multiple photon absorption events within a single
pulse.


\begin{thebibliography}{99}
\bibitem{footnote:1}
Photon antibunching can also be observed when the coupled
emitter-cavity system is not in the strong coupling regime.
\end{thebibliography}
\begin{thebibliography}{99}
\bibitem{footnote:S1}
When comparing polariton and cavity lifetimes in Fig. S~5b, we would
expect efficient emission from the uncoupled cavity already at
$5~\mathrm{pW}$. However, cavity emission is strongly suppressed,
since an initial polariton excitation is required to bring the QD
into a charged/dark state.
\end{thebibliography}
\end{document}